# Key Parameters in Determining the Reactivity of Lithium Metal Battery

Bingyu Lu[1], Diyi Cheng[2], Bhagath Sreenarayanan[1], Weikang Li[1], Bhargav Bhamwala[1], Wurigumula Bao[3*], Ying Shirley Meng[1,3*]

[1]Department of NanoEngineering, University of California, San Diego, La Jolla, CA 92093, USA
[2]Materials Science and Engineering Program, University of California, San Diego, La Jolla, CA 92093, USA
[3]Pritzker School of Molecular Engineering, University of Chicago, Chicago, IL 60637

*Correspondence to: shirleymeng@uchicago.edu, wubao@uchicago.edu.

**Abstract:** Lithium (Li) metal anodes are essential for developing next-generation high-energy-density batteries. Numerous concerns on the potential safety hazards of the Li metal have been brought up before its massive application in commercialized battery packs. However, few investigations have been performed to systematically evaluate the reactivity of Li metal anode in full cell level. Here, differential scanning calorimetry (DSC) with in situ Fourier-transform infrared spectroscopy (FTIR) are used to quantitatively investigate the Li metal reactivity. Lithiated graphite (Li-Gr) and lithiated silicon (Li-Si) are also studied as the comparison samples. The reactivity of the plated Li when coupled with different electrolyte composition, morphology, and atmosphere is systematically studied. More importantly, the reactivity of Li metal full cell with different cathode materials (NMC622, LFP and LNMO) has been compared. It was found that all cell components, including electrolyte composition, Li morphology, the control of inactive Li accumulation and cathode stability, are essential in controlling the reactivity of the plated Li. After optimizing these conditions, the Li metal full cell shows no significant thermal reaction up to 400˚C. This work identifies the key parameters in controlling the reactivity of the plated Li and may facilitate lithium metal battery design and manufacturing in the coming future.

**Main text:** With the rapid growth in the demand of high performance electric vehicles and personal portable devices, lithium (Li) metal has been a popular candidate as the anode material for developing high energy density rechargeable batteries (> 500 Wh/kg) due to its high specific capacity (3,860 mAh/g) and low electrochemical potential (–3.04 V versus the standard hydrogen electrode)[1–3]. Although extensive studies have been performed to prolong the cycle life of Li metal anode[4–6], the potential safety hazard brought by metallic Li in the high energy density batteries is still one of the biggest obstacles before its commercialization with large-scale[7]. The first attempt to commercialize Li metal cells in the 1980s ended up as a failure when multiple cases of cells catching on fire were reported[8]. Since then, the safety concerns of using Li metal as the anode material for high energy cells have never ceased[9].

The safety hazards at cell level is determined by the reactivity of each cell component and the interplay among them.[10] As for commercial Li-ion battery, there has been tremendous work studying the key factors in determining the reactivity of Li-Gr[11–13]. By utilizing accelerating rate calorimetry (ARC) and X-ray diffraction (XRD), Dahn et al discovered that the lithiated Graphite ($Li_{0.81}C_6$) starts to decompose with electrolyte at temperatures as low as 90°C[14]. In situ synchrotron XRD and mass spectrometry (MS) were applied by Amine et al to study the role of robust SEI in protecting lithiated graphite from thermal decomposition. Based on the current literature, it can be concluded that the lithiated graphite (Li-Gr) at material level is far from a safe and stable material, but can be implemented in state-of-the-art battery packs with proper engineering and optimizations[15–17]. As a highly reactive alkali metal, Li has always been considered unsafe for practical battery operations[18,19]. Various attempts have been made to design a safe rechargeable Li metal cell. Novel electrolyte with non-flammable solvents is one of the most effective ways to prevent the cell from catching on fire[20–22]. Fire-retarding localized high concentration electrolytes (LHCEs) have also been developed using non-flammable solvents or diluents, such as trimethyl phosphate or 2,2,2-trifluoroethyl-1,1,2,2-tetrafluoroethyl ether (HFE)[23,24]. Recently, Yin and his colleagues developed a new type of liquefied gas electrolyte with fire extinguishing merit[22]. With the development of electrolyte, the Li metal anode is marching towards a commercial reality with safe operation.

Although these works have been done focusing on preventing cells from catching on fire, it is still unclear how reactive Li metal is in nature, not to mention a direct comparison with other anode materials under similar state of charge. It is commonly believed that the pristine Graphite and Si are much safer than Li metal when assembled in a cell. However, in practical applications, the battery is mainly in the charge state in which the anode is lithiated, and its chemical stability is reduced significantly. Therefore, the reactivity of Li metal should be compared with that of Li-Gr and Li-Si, instead of the pristine ones. More work needs to be done to identify the key parameters in controlling the reactivity of Li metal in a battery system and provide designing principles for safe Li metal battery (LMB). There are multiple ways to define the metal reactivity under different cirtumstances.[25] For instance, a metal is considered highly reactive when 1) it causes large negative enthalpy of formation, $\Delta H_f$, during a oxidation reaction, or 2) requires small sublimation energy and ionization energy during oxidation or hydration[25]. Since the anode at charged state is full of electrons to be released, which can also react with water and oxygen violently[6], it is important to *quantitatively* compare the reactivity of different anode materials in a well-controlled cell system.

Here, we quantitatively compare the reactivity of Li-Gr, Li-Si, and plated Li metal (plated-Li) in two different electrolytes by utilizing the differential scanning calorimetry (DSC) coupled with in situ Fourier-transform infrared spectroscopy (FTIR). We further explored the effect of Li morphology on the reactivity by precisely tuning the external stack pressure during the plating process. It was found that the thermal response of plated-Li metal in a well-designed system can be on the same magnitude as that of Li-Gr and Li-Si. Furthermore, the influence from the cathode on the Li metal reactivity is also analyzed. Finally, a guideline for designing safer Li metal cells is provided.

Three anodes including Gr, Si and bare Cu (no excess Li) were charged with the controlled lithiation/plating amount of 3 mAh/cm$^2$ for Gr or 5 mAh/cm$^2$ for Si and bare Cu in the half cell, as shown in Fig. 1. Instead of 5 mAh/cm$^2$, 3 mAh/cm$^2$ capacity is chosen for the Gr because it is the most widely available capacity among commercialized Li-ion cells. The prepared anodes were sealed in a DSC pan with controlled amount of electrolyte (E/C ratio ~ 3mg/mAh), and then transferred into the DSD-FTIR station for thermal analysis. Detailed experimental designs can be found in Supplementary information. All DSC tests are done at least twice to make sure that the obtained results are repeatable (Fig. S5). Fig. 2 shows the DSC curves of Li-Gr, Li-Si, plated-Li in the carbonate-based electrolyte (Carbonate, 1.2 M Lithium hexafluorophosphate (LiPF$_6$) dissolved in ethylene carbonate (EC): diethyl carbonate (DEC) (1:1 by weight) with 10% fluoroethylene carbonate (FEC)). In addition, plated-Li in LHCE (Lithium bis(fluorosulfonyl)imide (LiFSI), 1,2-dimethoxyethane (DME) and 1,1,2,2-Tetrafluoroethyl-2,2,3,3-Tetrafluoropropyl Ether (TTE) with molar ratio 1:1.2:3) was also prepared. In both Li-Gr and Li-Si, most heat-absorbing peaks are associated with the evaporation of electrolyte solvents such as DEC and EC (Fig. 2a, b, Table S1). No significant heat-releasing peaks exist in the Li-Gr and Li-Si samples when heated up to 400°C. However, when the plated-Li in the Carbonate is heated during the DSC measurement, two heat-releasing peaks overlap with the evaporations of DEC and EC solvents, respectively (Fig. 2c). When Li melted at around 180°C, a sharp heat-absorbing peak appeared. The exothermic reactions might be caused by the melted Li quickly reacting with the remaining EC solvent and LiPF$_6$ salt. Fig. 2d shows the DSC curves of the plated-Li in LHCE electrolyte. The DME and TTE solvents are mostly evaporated before 100°C because of the low evaporation points, rather than reacting with the Li. A sharp Li melting peak is also shown around 180°C in Fig. 2d, which indicates that the Li was mostly melted rather than oxidized during the heating process. A small oxidation peak can be observed after the complete melting of Li, which is associated with the decomposition of electrolyte salts (details in Fig. S2).

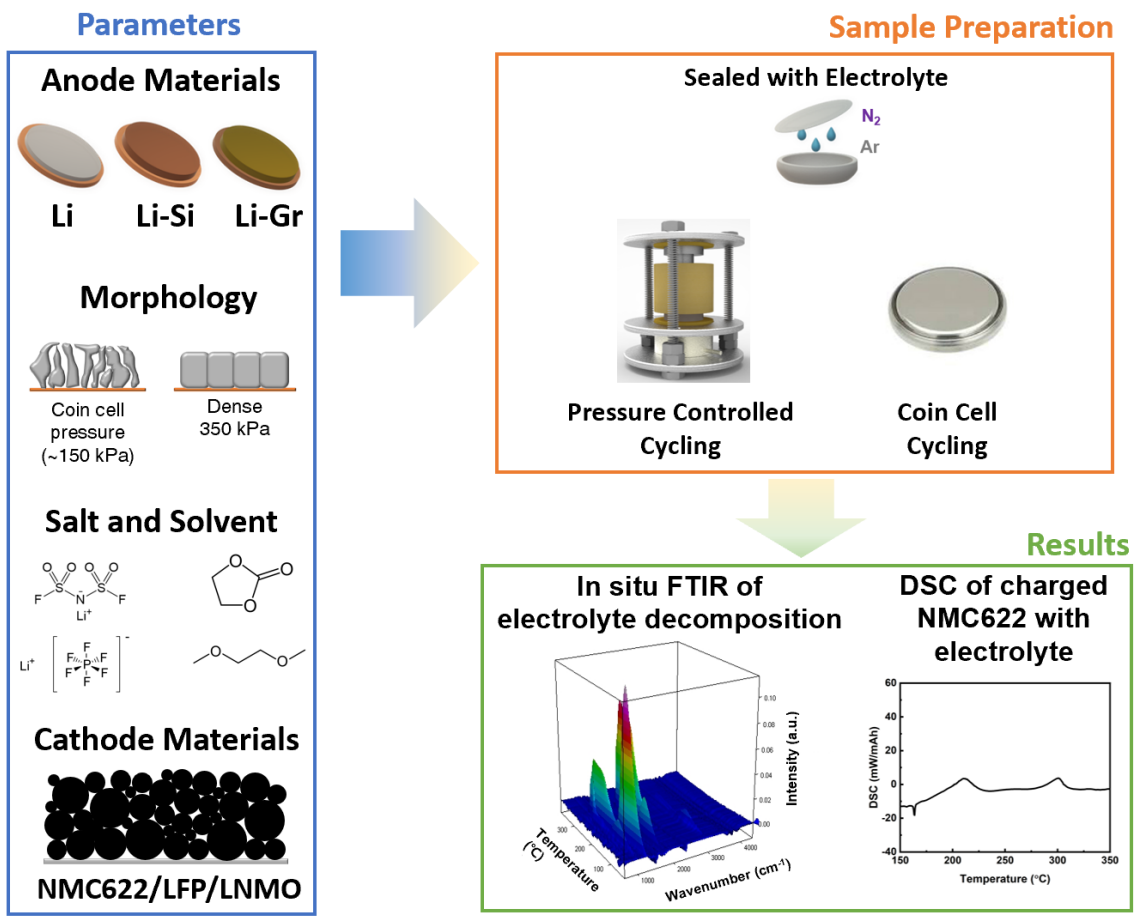

**Figure 1.** The schematics of the sample preparation and experimental process.

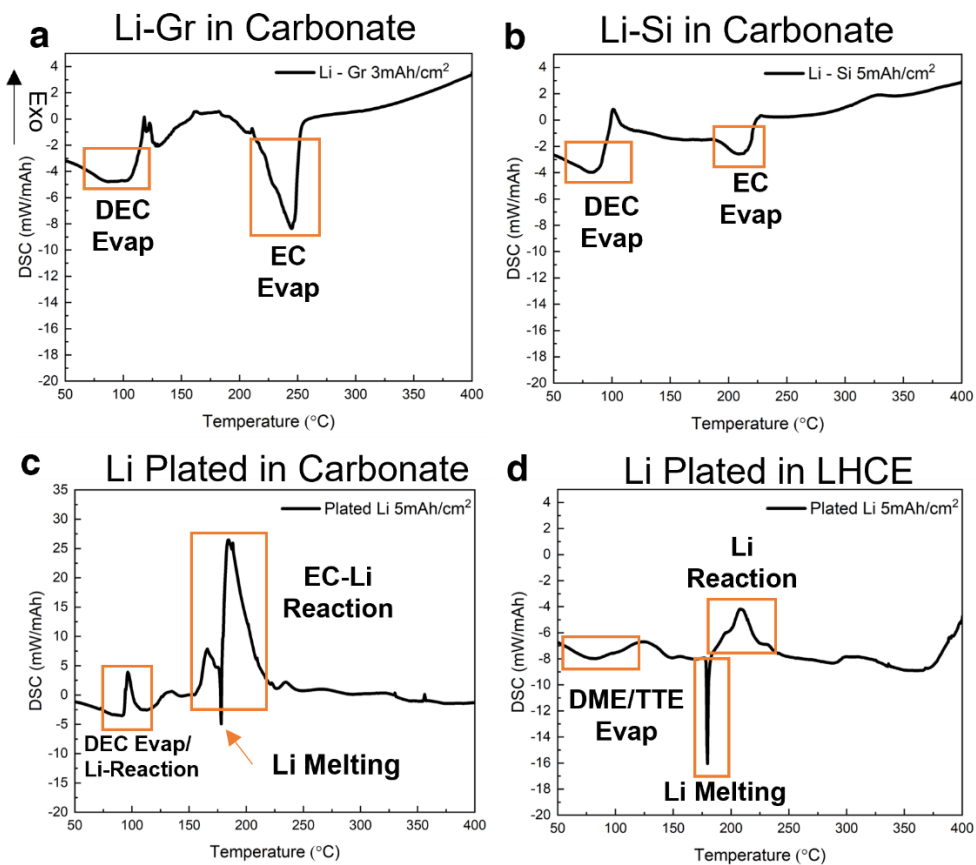

**Figure 2.** The DSC curves of (a) Li-Gr, (b) Li-Si, (c) Li metal plated in Carbonate and (d) Li metal plated in LHCE. Graphite and Si anodes are lithiated to the desired capacity in half cell setup with the rate of C/20. Li metal is plated in Li||Cu coin cell to the desired capacity at a current density of 0.5mA/cm$^2$.

The reactivity of the three types of anodes is also studied after 10 cycles. Fig. 3a-c shows the DSC curve and the morphology of the lithiated Graphite anode after 10 cycles. Similar to the 1$^{st}$ cycle charged Graphite samples, most of the DSC peaks are associated with the evaporation of the electrolyte solvents, with the exception of two exothermic peaks during the solvent evaporation. As shown in Fig. 3b and 3c, there seems to be some SEI accumulation on the Graphite surface as the layered surface morphology of graphite has disappeared after 10 cycles. The small heat releasing peaks (at around 100°C and 230°C) might be associated with the oxidation of the Li-Gr electrode during the solvent evaporation process. Similar trend is found in the 10-cycled Si anode (Fig. 3d-f), where two exothermic peaks (at around 100°C and 230°C) are also found during the same temperature range. The SEI accumulation is also obvious on the Si surface (Fig. 3e-f). The accumulation of SEI and trapped Li in the Gr and Si electrode after 10 cycles might contribute to the two small heat releasing peaks observed in the DSC. Overall, the reactivity of lithiated Si and Graphite is relatively low as no large exothermic peaks are observed during the heating process. Fig. 3g shows the DSC curve of the plated-Li in LHCE electrolyte after 10 plate-strip cycles. As the inserted images and SEM images are shown (Fig. 3h-i), because of the superior performance of the LHCE, the deposited Li is still shiny (Fig. S1), and the Li particles are bulky after 10 cycles.

In addition to that, most of the electrolyte solvents are evaporated before the Li melting point. As a result, the thermal response of the Li in LHCE is still relatively low, with only 39.8J/mAh of heat released during the heating process, which is in the same magnitude as the Li-Gr and Li-Si cases. However, for the plated-Li in Carbonate, there is a large amount of mossy Li accumulated on the electrode surface (Fig. 3k-l and insert of 3j). Because of the low cycling Coulombic efficiency (CE) of the Carbonate, there is a significant amount of nano-size inactive Li accumulated on the electrode, which can be seen as the mossy Li (Fig. 3l)[26]. The accumulation of these nano-size inactive Li eventually caused an explosion of the DSC Pan during the heating process. Based on the results so far, it can be seen that both the electrolyte and the morphology of Li play significant roles in controlling the reactivity of Li.

In situ FTIR is used to decipher the gas evolution during the DSC of Li metal to study the effects of electrolyte on the plated-Li. Fig. 4a shows the FTIR spectra of the gas generated during the heating process of plated-Li with Carbonate. Before 130°C, most of the peaks are associated with the evaporation of DEC (boiling point: 127°C). The spectra have two main peaks located at 1268 cm$^{-1}$ and 1771 cm$^{-1}$, representing O-C-O symmetric stretch and C=O stretch on the DEC molecules, respectively[27]. EC's evaporation occurs at higher temperatures because of the higher boiling point of 243°C. The peaks associated with EC are located at 1090 cm$^{-1}$ and 1875 cm$^{-1}$, representing O-C-O asymmetric stretch and C=O stretch on the EC molecules respectively[27]. There is no other obvious peak detected during the in situ FTIR study for the plated-Li with Carbonate samples. The overlapping temperature ranges between Li melting and solvent boiling can be part of the reasons why the Li show large exothermic peak during the DSC measurement (Fig. 2c).

Contrary to the Carbonate, because of the low boiling point of TTE (93.2°C) and DME (85°C), the solvents in the LHCE electrolyte evaporated before 120°C (Fig. 4b). As a result, the oxidation of Li is caused mainly by the residue organic components and electrolyte salts, as shown by N=O and $C_xH_yF_z^+$ fragment peaks in the FTIR spectrum at 200°C. The DSC of the pristine LHCE also confirms this observation (Fig. S2). The results so far confirm that the electrolyte solvents are the main reasons for the thermal instability of Li metal cell. However, the future of electrolyte design not only should study the fire-retarding features of the solvents, but also needs to focus on the thermal stability of the electrolyte salts, as the decomposition of electrolyte salt, which leads to the breakage of the S-N bond in the FSI anion to form a FSO2N· radical[28], will also oxidize the molten Li, interact with the delithiated cathode, eventually causes the release of heat or even fire[29].

The effects of Li morphology on the Li reactivity are also studied. A split cell setup (Fig. 5a) is used to control the morphology of the plated Li by tuning the external stacking pressure[30]. As shown in Fig. 5b-c, even with the Carbonate electrolyte, which is known for producing Li whiskers, the plated Li can achieve nearly 100% dense morphology. With the improved morphology, the Li porosity is significantly reduced, as shown in Fig. 5e. The plated-Li shows a relatively slow oxidation process instead of a sudden heat release as in Fig. 5d. Based on the results, it can be concluded that both the electrolyte and Li morphology play crucial roles in controlling the Li reactivity. Most of the oxidation of Li takes place between Li and electrolyte. If the Li can be plated in a nearly 100% dense morphology, the contact surface area between Li and the electrolyte can be significantly reduced so that the oxidation of Li metal by the electrolyte can also be largely slowed down.

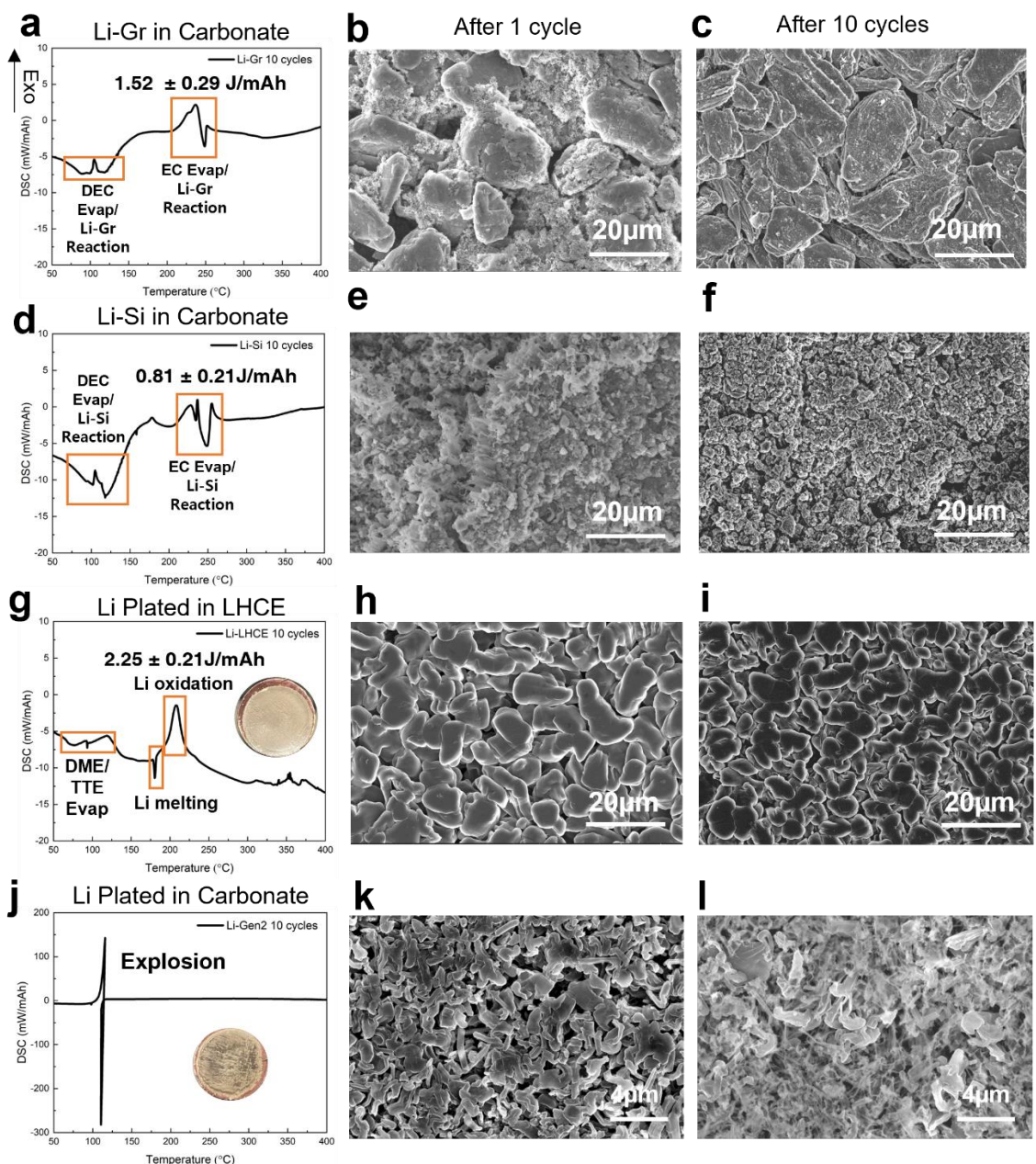

**Figure 3.** The DSC curves of (a) Li-Gr, (d) Li-Si, (g) Li plated in LHCE and (j) Li plated in Carbonate after 10 cycles. The SEM images of anode morphology after 1 cycle and after 10 cycles: (b-c) graphite, (e-f) Si, (h-i) Li plated in LHCE and (k-l) Li plated in Carbonated. Inserts: the digital images of Li plated on the Cu after 10 cycles. The amount of heat release from the oxidation peak of each DSC curve is labeled in the corresponding figures. Graphite and Si anodes are cycled in half cell configuration at rate of C/20 and Li metal anodes are cycled in Li||Cu cells at rate of 0.5mA/cm$^2$.

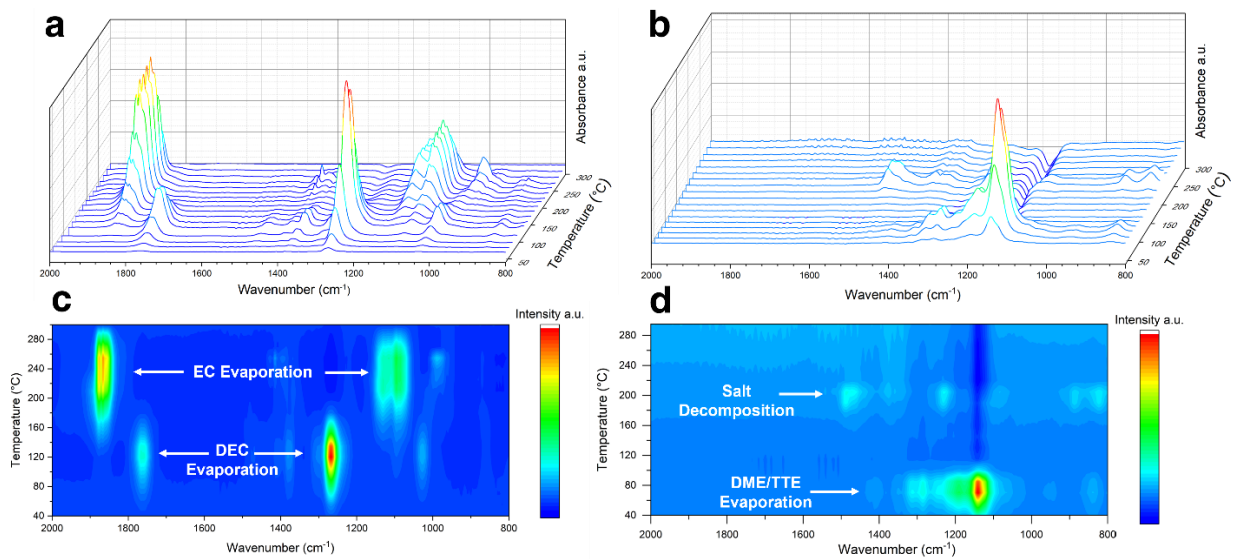

**Figure 4**. The in situ FTIR spectra of (a) plated Li with Carbonate and (b) plated Li with LHCE at different temperatures. The 2D intensity mapping of the in situ FTIR spectra of (a) plated Li with Carbonate and (b) plated Li with LHCE at different temperatures

The reactivity of Li metal full cell is also analyzed. With the knowledge gained so far, we utilized split cell setup to cycle the Cu||NMC622 full cell with LHCE to achieve uniform Li morphology. After the charging cycle, the cell is dissembled at 4.4V, and all the cell components are sealed into an pan for DSC measurement. Fig. 6a shows the DSC curve of delithiated NMC622 with electrolyte and separator. The delithiated NMC622 will decompose at around 220°C and 300°C to release $O_2$, which agrees with the literature results[31]. In addition to that, the decomposition of LiFSI salt is also observed at around 220°C, which overlaps with the decomposition of NMC622. When the delithiated NMC622 is coupled with plated Li, the released oxygen from the decomposed cathode and LiFSI salt will react with Li violently and cause a huge amount of heat release (Fig. 6b). Even if the Li morphology and the electrolyte are optimized, the $O_2$ released from the cathode is still detrimental to the full cell level safety. To further evaluate the impact of $O_2$ on the anode safety, DSC of Li-Gr, Li-Si and plated Li is done in air (Fig. S4). It was found that even though there is limited heat release from Li-Gr at lower temperature range (lower than 400°C), the carbon material started to burn in air because of the presence of $O_2$. The plated Li shows limited heat release because oxidation of Li in air was rather a slow process. Therefore, no sudden release of heat was observed (Fig. S4c). To minimize the impact of cathode on the Li metal full cell safety, we adopted $LiFePO_4$ (LFP) as the cathode material because of its stability under high temperatures. Although there is still some oxidation of Li caused by the decomposition of electrolyte salts (Fig. S2), the reduced release of $O_2$ indeed helps to improve the overall safety of the Li||LFP full cell, which releases only half of the heat as the Li||NMC622 did (Fig. 6d). To further minimize the impact from salt decomposition and $O_2$ release from the cathode, $LiNi_{0.5}Mn_{1.5}O_4$ (LNMO) cathode with all fluorinated electrolyte (All-F electrolyte, 1M $LiFP_6$ in FEC: Methyl 2,2,2-Trifluoroethyl Carbonate (FEMC), 3:7 by weight) is adopted for this purpose. As shown in Fig. 6e-f, the ultra thermal stability of both the cathode and electrolyte lead to an extraordinarily stable Li metal full cell. There is no obvious release of heat from the DSC test of the Li‖LNMO full cell after charging to 4.85 V. The effect of surface coating on the cathode is also investigated in this work. An thin layer (~2nm) of $Al_2O_3$ is coated onto the NMC622 cathode through Atomic Layer Deposition (ALD) method[32]. With the applied coating, we hope to stop the interactions between the electrolyte salt and the delithiated cathode to mitigate the decomposition of the cathode[29]. In Fig. S6, the thermal stability of the Li‖coated NMC622 full cell is tested with LHCE. The DSC shows that even with the $Al_2O_3$ coating, the thermal stability of the full cell did not improve much. However, it is known so far that the LiFSI salt in the LHCE electrolyte will decompose at around 220˚C, which might contribute to this instability. Therefore, the more stable All-F electrolyte is used in the Li‖ coated NMC622 full to study whether the thermal stability of the NMC622 is truly improved or not. As shown in Fig. 6g-h, no significant thermal stability improvement is observed in the $Al_2O_3$ coated-NMC622 sample compared to the uncoated one. The intrinsic instability of delithiated NMC622 will cause the release of $O_2$ and lead to a catastrophic heat generation when coupled with Li.

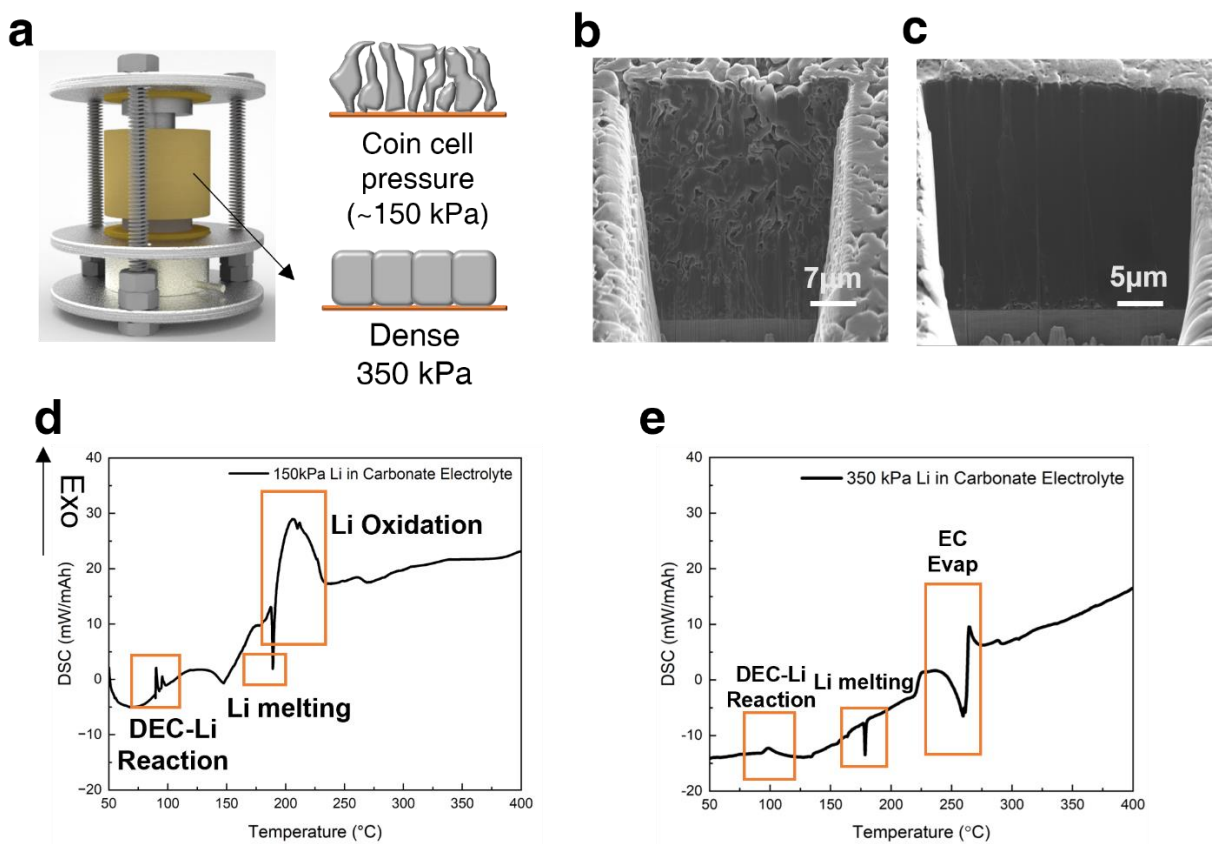

**Figure 5.** (a) The split cell setup for controlling the stack pressure on Li plating. The cross-sectional morphology of Li plated under (b) 150 kPa stack pressure (c) 350 kPa stack pressure. DSC curve of Li plated under (d) 150 kPa stack pressure and (e) 350 kPa stack pressure. All Li is plated to 5mAh/cm$^2$ at rate of 0.5mA/cm$^2$.

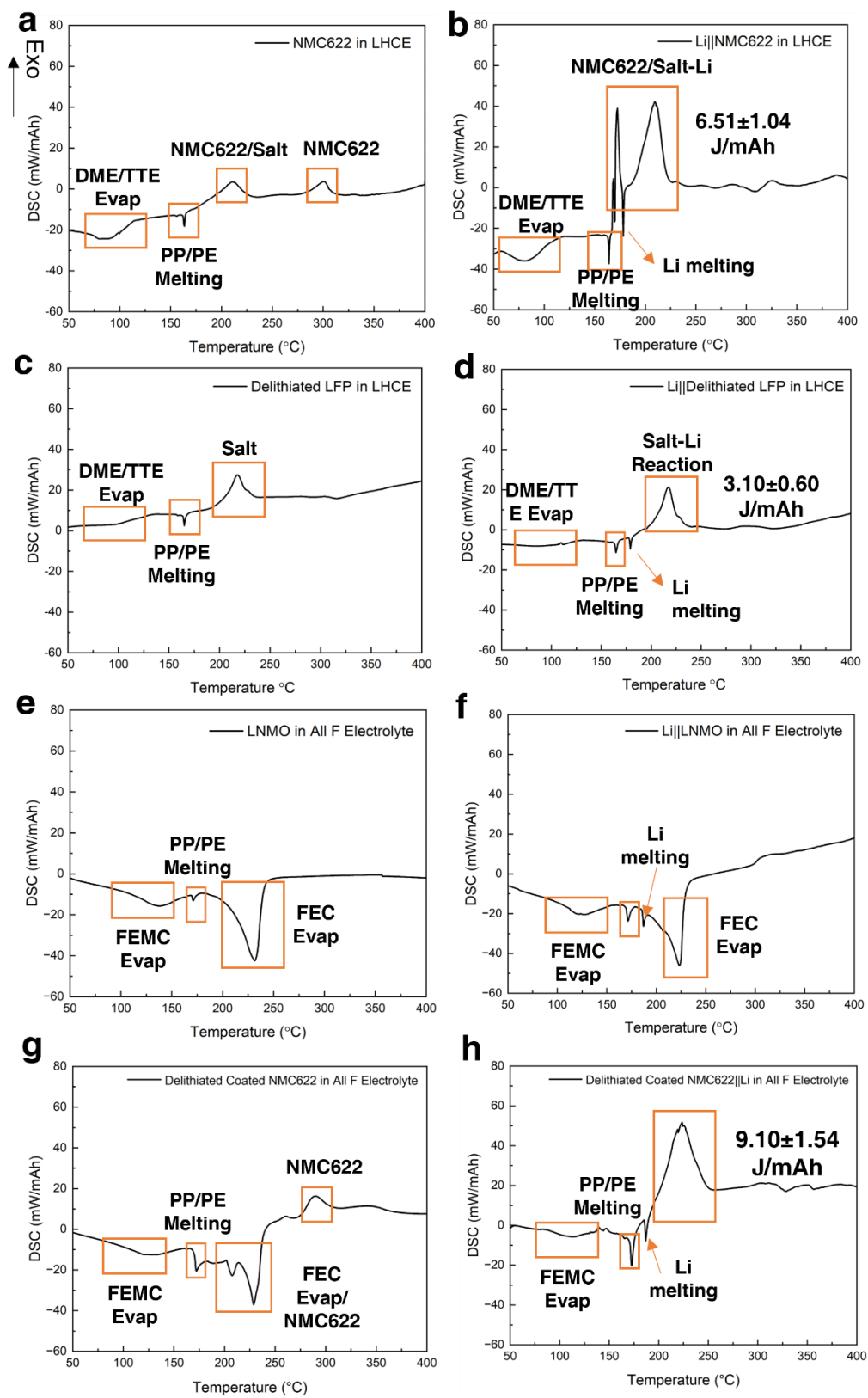

**Figure 6.** The DSC curves of (a) delithiated NMC622 with separator and LHCE, (b) plated Li with delithiated NMC622, separator and LHCE, (c) delithiated LFP with separator and LHCE, (d) plated Li with delithiated LFP, separator and LHCE, (e) delithiated LNMO with separator and All-F electrolyte, (f) plated Li, delithiated LNMO with separator and All-F electrolyte, (g) delithiated coated NMC622 with separator and All-F electrolyte and (h) plated Li, delithiated coated NMC622 with separator and All-F electrolyte. All cells are cycled at C/20 with corresponding electrolyte.

In conclusion, the reactivity of Li-Gr, Li-Si, and plated Li are compared using integrated DSC and in situ FTIR techniques. It is found that the reactivity of plated Li in the cell is highly related to its morphology and the electrolyte composition. With dense morphology and novel electrolyte, the reactivity of plated Li in the cell can be drastically suppressed to the same level as that of Li-Gr and Li-Si anodes. Therefore, it is crucial to plate Li in dense morphology to minimize its surface area and utilize thermal-stable electrolytes for safe operation of Li metal cells. Moreover, the crosstalk influence from the cathode thermal decomposition may cause a safety hazard when Li metal anode is used. By switching to more thermally stable cathode materials such as LFP and LNMO, the thermal stability of the Li metal full cell can be largely improved. In addition to that, the decomposition of the electrolyte salt also needs to be strictly controlled. Lastly, the cycle number and cell environment all contribute to the Li metal reactivity. It is important to control the accumulation of inactive Li and Li morphology even after extended cycles. The key parameters in controlling the reactivity of Li metal discovered in this work can be applied to the future research of Li metal anode for practical Li metal full cells.

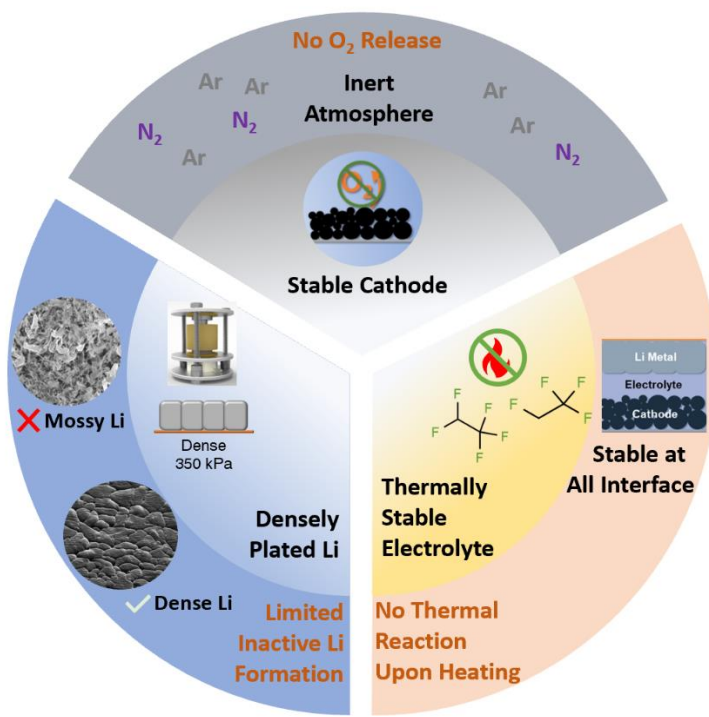

**Figure 7.** The crucial parameters that control the reactivity of Li metal cells: the outer ring represents the optimized conditions for Li metal anode while the inner ring shows the unoptimized conditions.

**Conflicts of interest:**
There are no conflicts of interest to declare among the authors.

**Acknowledgments:**

**Funding:** This work was supported by the Office of Vehicle Technologies of the U.S. Department of Energy through the Advanced Battery Materials Research (BMR) Program (Battery500 Consortium) under Contract DE-EE0007764. Cryo-FIB was performed at the San Diego Nanotechnology Infrastructure (SDNI), a member of the National Nanotechnology Coordinated Infrastructure, which is supported by the National Science Foundation (grant ECCS-1542148). We acknowledge the UC Irvine Materials Research Institute (IMRI) for the use of the DSC-FTIR, funded in part by the National Science Foundation Major Research Instrumentation Program under Grant CHE-1338173. The authors would like to acknowledge Neware Technology Limited for the donation of BTS4000 cyclers which are used for testing the cells in this work.

**Method**

**Electrolyte**. Battery-grade lithium bis(fluorosulfonyl)imide (LiFSI) was purchased from Oakwood Products, Inc.; Bis(trifluoromethane)sulfonimide lithium 99.95% (LiTFSI) was purchased from Sigma-Aldrich. All salts were further dried at 120 °C under vacuum for 24 h before use; 1,2-dimethoxyethane (DME) anhydrous, 99.5% was purchased from Sigma-Aldrich; 1,1,2,2-tetrafluoroethyl-2,2,3,3-tetrafluoropropyl ether (TTE, 99%) was ordered from SynQuest Laboratories. Solvents were dried with molecular sieves before use. LiFSI–DME–TTE were mixed in a molar ratio of 1:1.2:3 to prepare the LHCE. The carbonate electrolyte, 1.2 M Lithium hexafluorophosphate (LiPF6) dissolved in ethylene carbonate (EC): diethyl carbonate (DEC) (1:1 by weight) with 10% fluoroethylene carbonate (FEC) was purchased from Gotion. The All-F electrolyte was directly made by Gotion with formula provided by UCSD. All procedures were performed in an argon gas filled glove box (<0.1 ppm $O_2$, <0.1 ppm $H_2O$).

**Electrochemical Testing.** For Li‖Cu cells, the cleaned Cu pieces was assembled in the 2032 coin cell as the working electrode while the Li metal (0.1 mm thick, China Energy Lithium Co., Ltd.) was the reference and counter electrode. Celgard 2325 separator was used as the separator and soaked in 55 µL of electrolyte. The Graphite and Silicon half cell were cycled at room temperature at a rate of C/20 during the first cycle and C/10 for subsequent cycles. The Graphite half cells are cycled between 0.05V to 2V while the Silicon half cells are cycled between 0.05V and 1.5V. For the full cell testing, the Graphite and Silicon electrode was paired with an LFP, NMC622, coated NMC622 or LNMO cathode and assembled in a 2032 type coin cell. The full cells were charged at room temperature at a rate of C/10 to 3.8V for LFP, 4.4V for NMC622 and coated NMC622 and 4.85V for LNMO. All cell makings were performed in an argon gas filled glove box (<0.1 ppm $O_2$, <0.1 ppm $H_2O$).

**Pressure controlled split cells.** A custom-made split cell that consists of two titanium plungers (1/2-inch diameter) and one polyether ether ketone (PEEK) die mold (1/2-inch inner diameter) is used for the pressure controlled Li plating. The Cu‖Li cells were made by sandwiching the Li metal foil (7 mm diameter, 50 µm thick, China Energy Lithium Co., Ltd.), Celgard 2325 separator (1/2 inch diameter) and the cleaned Cu foil between the two titanium plungers inside the PEEK die mold. Only minimum amount of electrolyte (~5 µL) was added to the Cu‖Li cells to wet the separator. After the assembly, the split cell and the load cell were put into the cell holder, which provided the uniaxial stacking pressure. The uniaxial stacking pressure was adjusted by the three screws on the cell holder. The screws were carefully adjusted to apply the desired stacking pressure to the split cell while keeping both the split cell and the load cell in vertical position. The cell was tested inside the glovebox using Landt CT2001A battery cycler (Wuhan, China). Various current densities and stacking pressure were applied to conduct the study as indicated in the main text.

**Cryogenic Focused Ion Beam- Scanning electron Microscopy (Cryo-FIB/SEM)**. The copper foil with deposited Li was recovered from the split cell and then washed with DME to remove the residual electrolyte in the Ar-filled glovebox. The sample was mounted on a SEM stub (Ted Pella) in the glovebox, then transferred to a FEI Scios DualBeam FIB/SEM system with an Air-tight transfer holder to minimize air exposure[2]. Liquid $N_2$ was used to cooled down the sample stage to -180°C to create a cryogenic environment which helps minimize beam damage to the sample. Gallium ion beam with a voltage of 30 kV, current of 7 nA and dwell time of 100 ns was used to roughly mill down the cross-section of the deposited lithium. After the rough milling, the cross-section was cleaned with ion beam at 1 nA. The SEM image of the cross-section was taken using Everhart-Thornley Detector (ETD) at 5 kV and 0.1 nA.

**ALD coating for NMC.** For the ALD process, the calendared cathode was first stored in a 60 ℃ oven overnight to remove the moisture, then transferred to the ALD chamber. The deposition of $Al_2O_3$ requires trimethylaluminum (TMA) as the precursor and water as the reactor. The carrier gas was nitrogen in 300 mbar, and the reaction temperature was 100 °C. The deposition rate was 1.0 Å per cycle. The surface layer thickness on the electrodes was controlled through the number of cycles performed. To ensure the precursor gas could spread into the electrode, a pre-injection of TMA for 6 seconds was applied. In the rest of the cycles, the TMA dose time was 0.6 s, followed by 1 s TMA purge, and the moisture dose time was 0.2 s, followed by 1 s purge. A vacuum-drying process at 80 °C for at least 24 h was then applied to the surface-modified electrode before any electrochemical testing to remove any residual moisture.

**Differential scanning calorimetry - Fourier-transform infrared spectroscopy (DSC-FTIR).** The DSC-FTIR measurement was done on NETZSCH STA 449 F3 Jupiter with an in-line coupled system of Bruker ALPHA II FTIR. The cycled electrodes were first retrieved from the coin cells and seal in a Cu backed Al pan with controlled amount of electrolyte (~3 g/Ah). All sample preparations were performed in an argon gas filled glove box (<0.1 ppm $O_2$, <0.1 ppm $H_2O$). The Al pan was pierced by a needle after loading into the DSC chamber so that the evolved gas could be analyzed by the in-line FTIR. During the DSC-FITR measurement, the temperature was ramping up at a rate of 10°C/min to 400°C. All DSC-FTIR measurement were done under Ar and $N_2$ environment.

| Solvent | Boiling Point (°C) |
|---|---|
| DME | 85.0 |
| TTE | 93.2 |
| DEC | 127.0 |
| EC | 243.0 |
| FEC | 212.0 |
| FEMC | 90.0 |

**Table S1**. The boiling point of all the electrolyte solvents used in the study.

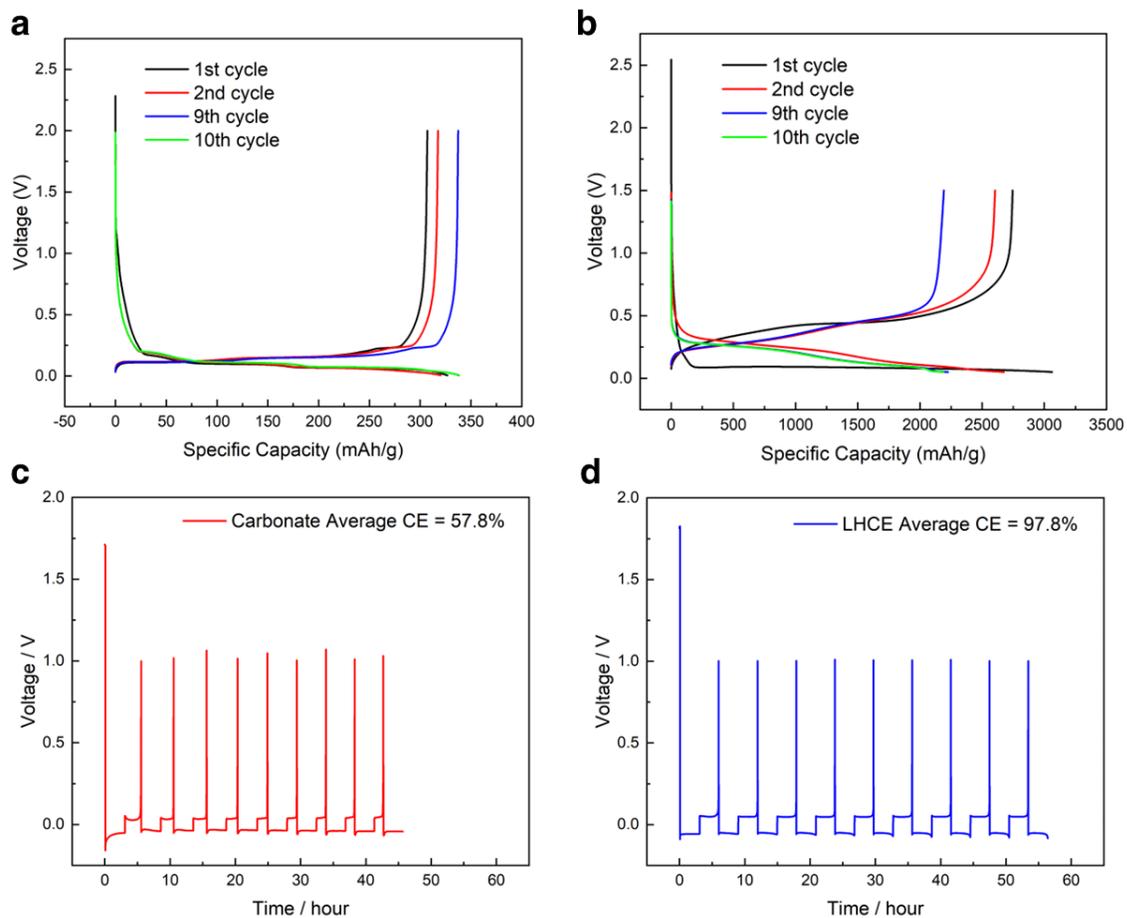

**Figure S1**. The cycling voltage profiles of (a)Graphite, (b)μSi, (c)Li||Cu in Carbonate and (d) Li||Cu in LHCE.

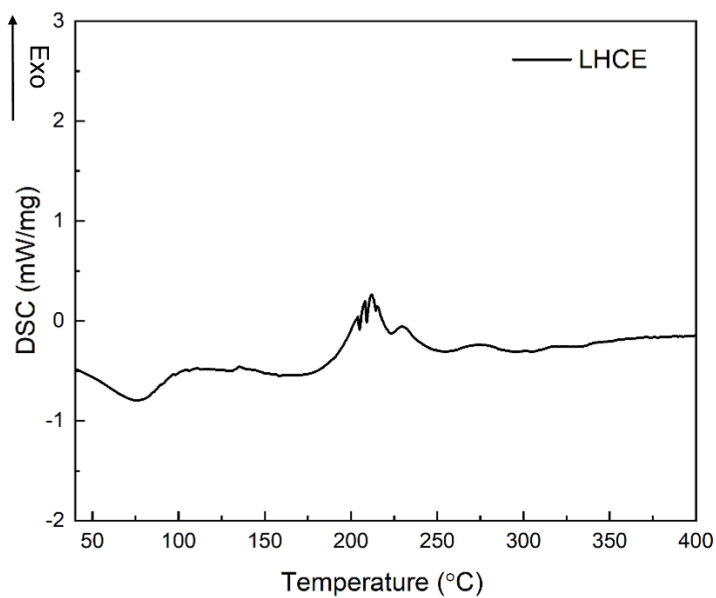

**Figure S2**. The DSC profile of LHCE.

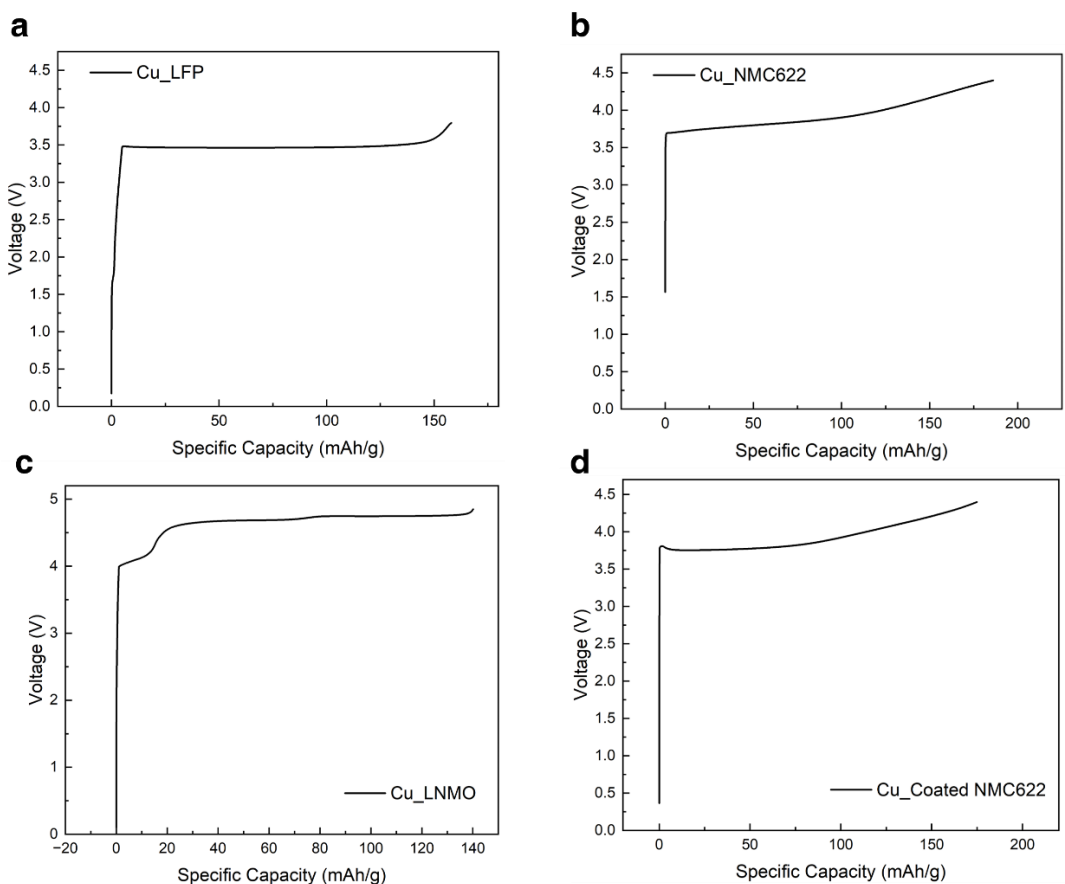

**Figure S3**. The cycling voltage profiles of (a) Cu||LFP, (b) Cu||NMC622, (c) Cu||LNMO and (d) Cu||Coated NMC622.

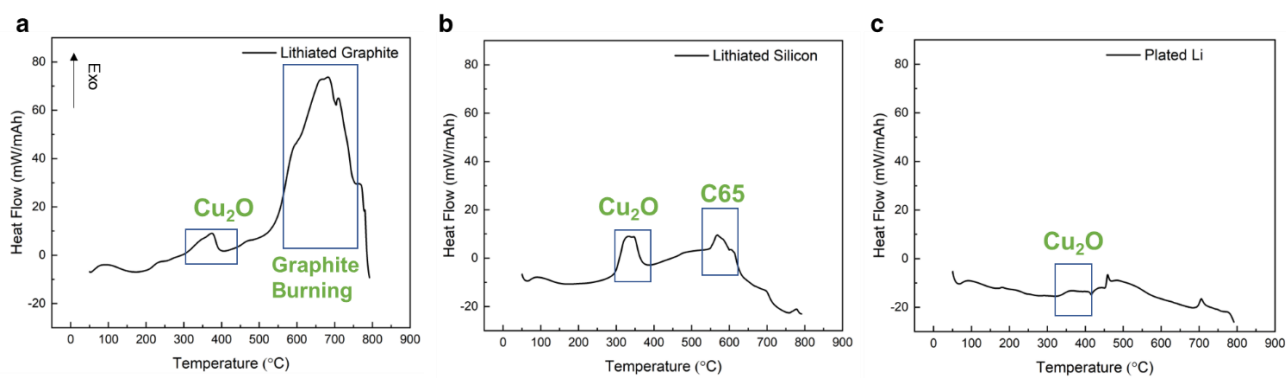

**Figure S4**. The DSC profiles of (a) Li-Gr (b) Li-Si and (c) plated-Li in Carbonate. All DSC is done in air. Graphite and Si anodes are cycled in half cell configuration at rate of C/20 and Li metal anodes are cycled in Li||Cu cells at rate of 0.5mA/cm$^2$. Carbonate electrolyte is used for all cells.

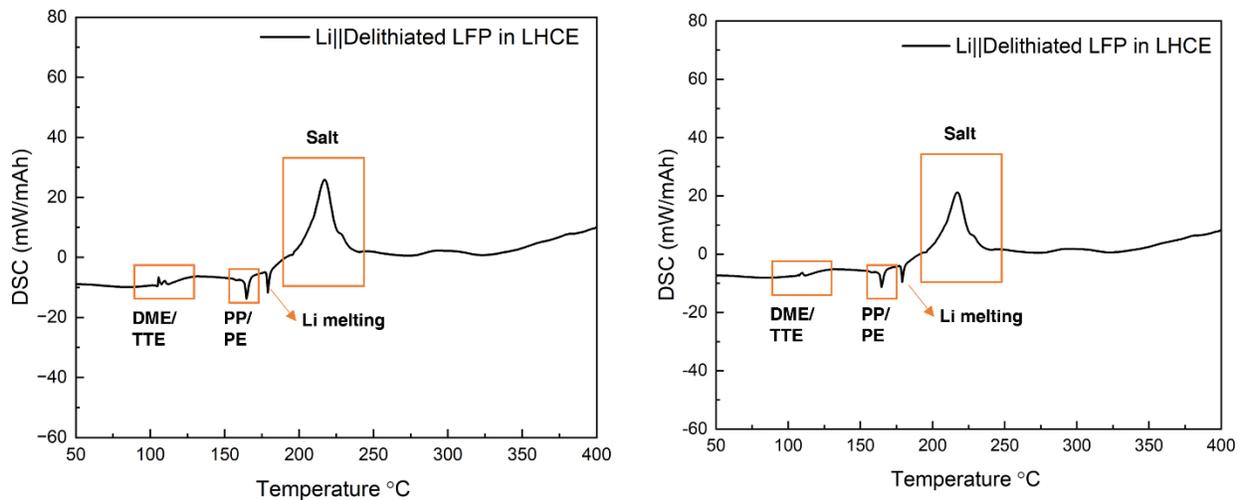

**Figure S5**. The DSC profiles of duplicated tests of delithiated LFP in LHCE to show the reproducibility of the DSC tests.

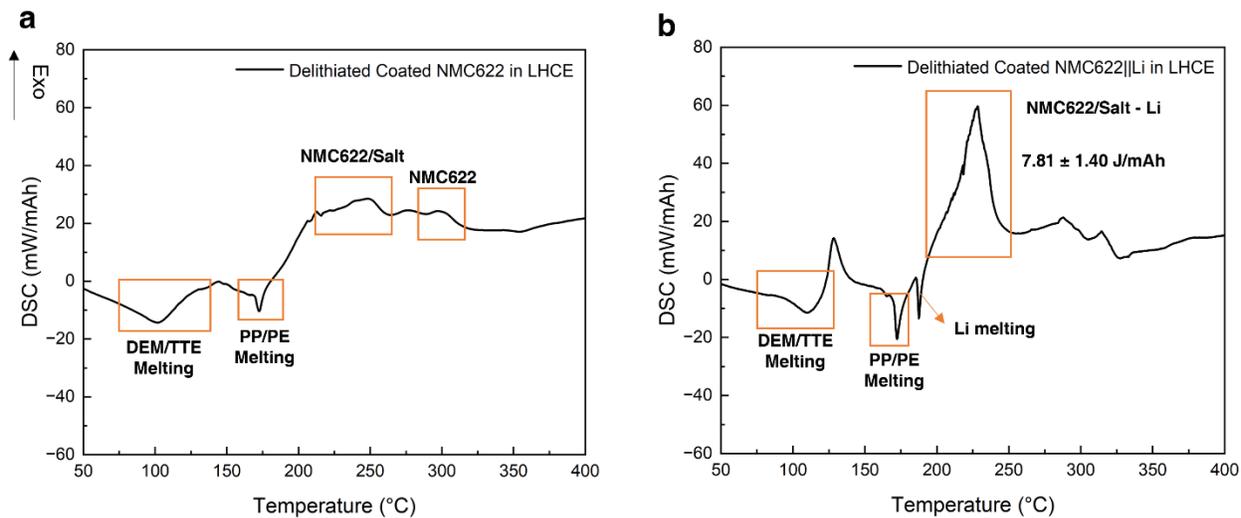

**Figure S6**. The DSC curves of (a) delithiated coated NMC622 with separator and LHCE and (b) plated Li, delithiated coated NMC622 with separator and LHCE. All cells are cycled at C/20 with corresponding electrolyte.